# Non-locally sensing the magnetic states of nanoscale antiferromagnets with an atomic spin sensor

Shichao Yan[1,2,†,*], Luigi Malavolti[1,2], Jacob A. J. Burgess[1,2], Andrea Droghetti[3], Angel Rubio[1,3], Sebastian Loth[1,2,4,*]

[1] *Max Planck Institute for the Structure and Dynamics of Matter, 22761 Hamburg, Germany*

[2] *Max Planck Institute for Solid State Research, 70569 Stuttgart, Germany*

[3] *Nano-Bio Spectroscopy Group, Department of Materials Science, Universidad del País Vasco , Avenida Tolosa 72, 20018 San Sebastian, Spain*

[4] *Institute for Functional Materials and Quantum Technology, University of Stuttgart, 70569 Stuttgart, Germany*

[†] *present address: Department of Physics, University of Illinois at Urbana-Champaign, IL 61801 USA*

*Email: sebastian.loth@mpsd.mpg.de, yansc@illinois.edu

**The ability to sense the magnetic state of individual magnetic nano-objects is a key capability for powerful applications ranging from readout of ultra-dense magnetic memory to the measurement of spins in complex structures with nanometer precision. Magnetic nano-objects require extremely sensitive sensors and detection methods. Here we create an atomic spin sensor consisting of three Fe atoms and show that it can detect nanoscale antiferromagnets through minute surface-mediated magnetic interaction. Coupling, even to an object with no net spin and having vanishing dipolar stray field, modifies the transition matrix element between two spin states of the Fe-atom-based spin sensor that changes the sensor's spin relaxation time. The sensor can detect nanoscale antiferromagnets at up to three nanometers distance and achieves an energy resolution of 10 micro-electronvolts surpassing the thermal limit of conventional scanning probe spectroscopy. This scheme permits simultaneous sensing of multiple antiferromagnets with a single spin sensor integrated onto the surface.**

# Introduction

Properties of quantum spin systems, such as magnetic stability (1) and spin coherence (2, 3), depend sensitively on their local conditions. Therefore, they can be used to great effect as sensors for magnetic environments. Great effort is directed toward increasing the spatial resolution and sensitivity of spin detection with a variety of techniques using quantum systems, such as magnetic resonance force microscopy (4, 5), scanning nano-SQUIDs (6), and nitrogen-vacancy (NV) centers in diamond (7-10) which have achieved nanometer resolution. A scanning tunneling microscope (STM) is capable of atomic resolution, but the direct interaction of the microscope tip can have an invasive influence on spins (11). Controllably placing the sensor in close proximity, ideally with atomic precision, is crucial and is an intensely studied problem for NV centers and other nanofabricated field sensors. Such atomic-scale precision is routine in STM-based atom manipulation, albeit at cryogenic temperatures and on monocrystalline surfaces (12, 13). Manipulation of magnetic adatoms allows the creation of quantum spin systems that can be addressed by single-atom magnetometry (14, 15), inelastic electron tunneling (16, 17), electronic pump-probe spectroscopy (1, 18), and electron paramagnetic resonance (19). Magnetic atoms on surfaces couple to their local environment via a broad range of interactions, for example by exchange (20), superexchange (21), non-collinear exchange (22), or RKKY (Ruderman-Kittel-Kasuya-Yosida) (15, 23) coupling; via Kondo scattering (24, 25); with electric (19) or magnetic fields; and to surface strain through magnetic anisotropy (17, 26). Adatom sensors therefore have the potential to realize detection schemes via any of these coupling mechanisms.

## Results and discussion

Here we show that measuring the spin dynamics of a few-atom spin system permits it to function as a highly sensitive surface-integrated sensor capable of detecting the presence and state of nearby magnetic nano-objects. Using a three Fe atom chain (trimer) we are able to detect switching of multiple nano-antiferromagnets via a substrate-mediated interaction, Fig. 1A. Both the trimer and the nano-antiferromagnets are assembled using a low temperature STM with Fe atoms on a monolayer copper nitride ($Cu_2N$) surface on a Cu(100) substrate (17, 25). The nano-antiferromagnets are constructed with an even number of antiferromagnetically coupled atoms so that any single one has two stable magnetic states, Néel states, with zero net spin. This is signified in spin-polarized STM topographs by alternating apparent height of the constituent Fe atoms (Fig. 1B). These states also exhibit spontaneous switching (13). The solely internal magnetic structure and absence of a net magnetic moment makes state switching in antiferromagnets difficult to detect by conventional magnetometry (5-7).

The trimer was chosen as the inaugural sensor because it features an avoided level crossing in its two low-energy spin states at zero magnetic field (20). The crossing mixes the $|+2\ -2\ +2\rangle$ and $|-2\ +2\ -2\rangle$ states of the trimer (±2 denotes the expectation value of each Fe atom's spin along the easy magnetic axis). This mixing permits transitions between the two states with a rate that is strongly dependent on magnetic field or any other local magnetic perturbation, Fig. 1C. A small magnetic field parallel to the $z$ axis, $B_0 > 0$, biases these two states so that the ground state, $|\phi\rangle$, becomes mostly $|+2\ -2\ +2\rangle$ (see Fig. S1) making it detectable by low bias voltage spin-polarized STM imaging, Fig. 1B (see Methods). Spin-polarized pump-probe measurements can then be used to pump the trimer to the exited state, $|\rho\rangle$, and probe the spin relaxation time, $T_1$, back to the ground state (18). Monitoring variations in $T_1$ reveals magnetic perturbations caused by changes in the trimer's local environment, Fig. 1B.

## Sensing the magnetic state of a nearby nano-antiferromagnet

With an external magnetic field of 0.25 T the Fe trimer has an average spin relaxation time, $T_1$, of 78 ns. We find that the Fe trimer switches between two different spin relaxation times as a nano-antiferromagnet constructed nearby switches between its two Néel states (see Supplementary Materials Section 1). The decay curve measured on the Fe trimer yields a longer $T_1$ when the nano-antiferromagnet is in Néel state "0" and shorter $T_1$ when it is in Néel state "1", Fig. 1B. The difference in spin relaxation time, $\Delta T_1$, is 26 ± 3ns measured for a separation of 3 nm between the Fe trimer and the nano-antiferromagnet. This demonstrates that Fe trimer can measure the magnetic state of a nearby antiferromagnet and act as a spin sensor.

This capability allows us to detect spontaneous switching of the nano-antiferromagnet by continuously monitoring the Fe trimer's spin relaxation time. When the nano-antiferromagnet switches from one Néel state to the other, the electronic pump-probe signal jumps between two decay curves, Fig. 1B. A trace of this two-state switching can be obtained by keeping the pump-probe delay fixed and measuring the pump-probe signal amplitude, Fig. 1D. Large pump pulse amplitudes induced additional switches in these traces (see Fig. S2). To minimize this effect we employed minimal amplitude pump pulses (10 mV). The amplitude of the two-state signal depends on the delay time chosen. Increasing the delay time from 150 ns to 250 ns reduces the signal and it vanishes when the Fe trimer's spin is fully relaxed at a delay of 500 ns, Fig. 1D. This further confirms the two-state switching is indeed encoded in the dynamic response of the Fe trimer.

The variation in $T_1$ ($\Delta T_1$) can be increased by reducing the separation of the trimer and the nano-antiferromagnet. The measurements investigating the Fe trimer-nano-antiferromagnet separation were performed with the same nano-antiferromagnet on the same copper nitride patch, simply moving the Fe trimer to different locations by STM atom manipulation (Fig. 2A). $\Delta T_1$ becomes larger as the separation between the nano-antiferromagnets and Fe trimer decreases (Fig. 2B): at 3.0 nm separation $\Delta T_1$ is 29 ns and increases to 466 ns at 1.1 nm separation.

**Exploring the sensing mechanism**

To understand the sensing mechanism, we inspect the spin relaxation process, which occurs predominantly by electron-spin scattering (18, 20). $T_1$ is determined by the product of the transition matrix element between the trimer's two low-energy spin states, $|\phi\rangle$ and $|\rho\rangle$, and the number of substrate electrons that can scatter (20) (see Supplementary Section 2). It is minimal at zero magnetic field, where transverse magneto-crystalline anisotropy induces the avoided level crossing between $|\phi\rangle$ and $|\rho\rangle$. Application of a magnetic bias field, $B_0$, parallel to the uniaxial anisotropy axis of the Fe atoms creates an energy splitting, $E_{\rho\phi}$, between the two low-energy states, reduces state mixing, and increases $T_1$ to a detectable level, Fig. 3A. This magnetic bias field sets a working point within the Fe trimer's two-level system. All other magnetic perturbations introduce an additional variation in the energy splitting, $\Delta E$, that leads to a variation in the spin relaxation time, $\Delta T_1$.

We use a spin Hamiltonian model to calculate the response of $T_1$ to different magnetic perturbations including longitudinal magnetic field, $B_\parallel$, transverse magnetic fields, $B_{\perp x,y}$, that add to the bias field ($B_0$) and a Heisenberg-type exchange interaction, $J_{nAF}$, with the nano-antiferromagnet, Fig. 3B (see Supplementary Section 2). We find that $T_1$ is very sensitive to $B_\parallel$ and $J_{nAF}$, but it is not sensitive to $B_{\perp x,y}$. This rules out the possibility that the changes in $T_1$ of the Fe trimer are induced by transverse dipolar magnetic fields from the nearby nano-antiferromagnet.

Notably, the variation of the spin relaxation time, $\Delta T_1$, with the perturbation energy, $\Delta E$, induced by $B_\parallel$ and $J_{nAF}$ is expected to be linear over a large range of perturbation magnitudes. We use an external superconducting vector magnet to apply the bias field and an additional perturbation field in the same direction, $B_0 + B_\parallel$. The perturbation energy of $B_\parallel$ is given by the Zeeman energy of the Fe trimer, $\Delta E = 8\mu_B B_\parallel$ (the magnetic moment of the Fe trimer is $4\mu_B$) so that we can plot $\Delta T_1$ as a function of perturbation energy, $\Delta E$, for each experimental configuration. We find that $\Delta T_1 / \Delta E (B_\parallel)$ is indeed linear, Fig. 3B.

The two curves, $\Delta T_1 / \Delta E (B_\parallel)$ and $\Delta T_1 / \Delta E (J_{nAF})$, are almost identical, Fig. 3B. This similarity permits us to use the external magnetic field as a quantitative reference to deduce the interaction energy between trimer and nearby nano-antiferromagnet from the observed changes in $T_1$ induced by Néel state switching (see Supplementary Materials Section 3 and Fig. S3). We verified that the choice of magnetic bias field ($B_0$) has no influence on the responsivity of the Fe trimer by finding that $\Delta T_1$ induced by Néel state switching is constant over a large range of bias fields for all nano-antiferromagnet-sensor arrangements we studied (Fig. 3C) even though the absolute magnitude of the Fe trimer's lifetime changes significantly with bias field (Fig. 1C). This also demonstrates that perturbations from magnetic fields and exchange interaction add linearly. It is worth noting that at large negative $\Delta E$ the variation of $T_1$ levels off and reverses when the perturbation is sufficiently large to compensate the bias field and cause $|\phi\rangle$ and $|\rho\rangle$ to cross, Fig. 3 B. In all measurements we kept the bias field large enough to avoid this region.

This makes it possible to deduce a simple empirical equation to quantitatively extract the magnetic interaction energy between the Fe trimer and the nano-antiferromagnet from measured values of $\Delta T_1$. As the nano-antiferromagnet is in Néel-state "1", the magnetic interaction with the Fe trimer will decrease the energy difference between $|\phi\rangle$ and $|\rho\rangle$ by $-\Delta E/2$, whereas it will increase by $\Delta E/2$ for Néel-state "0", Fig. 3a. Since $T_1$ is linearly dependent on both $B_\parallel$ and $J_{nAF}$, $\Delta T_1 / \Delta E\,(B_\parallel)$ and $\Delta T_1 / \Delta E\,(J_{nAF})$, are almost identical we can express magnetic interaction in terms of the Zeeman energy of $B_\parallel$ and obtain the empirical relation:

$$\Delta E = \frac{4g\mu_B}{\left(\frac{dT_1}{dB_\parallel}\right)} \Delta T_1 \qquad (1)$$

where $\mu_B$ is the Bohr magneton and g the Fe trimer's g-factor (g = 2). $\left(\frac{dT_1}{dB_\parallel}\right)$ is the responsivity of $T_1$ to longitudinal magnetic field and was obtained experimentally by varying the magnitude of $B_\parallel$. It ranges from 450 ns/T to 830 ns/T for Fe trimers in different locations on the $Cu_2N$ surface (see Fig. S3). Fig. 2C shows the extracted magnetic interaction as a function of the separation between the Fe spin sensor and nano-antiferromagnet. To remove possible influence from variations in the crystal-field environment of the sensor we performed the calibration of $\left(\frac{dT_1}{dB_\parallel}\right)$ for each sensor-nano-antiferromagnet distance. The magnetic interaction energy increases quickly from 12 μeV to 224 μeV as the separation decreases from 3.0 nm to 1.1 nm.

This interaction is surprisingly long-ranged when compared to the previously reported super-exchange interaction on the $Cu_2N$/Cu(100) surface (27,28). Dipolar magnetic interaction can be calculated from the known positions of all Fe atoms in the two structures but accounts only for ~5% of the measured interaction energy because the structures' antiferromagnetic order cancels most dipolar fields (black line in Fig. 2C). This suggests the presence of an indirect exchange interaction through the $Cu_2N/Cu(100)$ surface. Long-ranged indirect exchange interaction has thus far been found only for adatoms on metallic surfaces with a surface state that mediates oscillatory RKKY-type interaction (15, 23). This interaction is absent for adatoms on Cu(100) (29, 30), but the influence of the $Cu_2N$ decoupling layer has not been considered yet.

We therefore performed *ab-initio* all-electron density functional theory (DFT) calculations in order to elucidate the microscopic origin of this long-ranged indirect exchange interaction mechanism. Due to the long-range nature of the interaction we cannot treat large isolated nano-objects deposited on the surface in the our supercell calculation approach, consequently we modeled the experimental geometry by one Fe atom ladder (to simulate the nano-antiferromagnet) and one Fe atom chain (to simulate the Fe trimer) (see Methods and Supplementary Materials for details). The magnetic interaction energy between Fe ladder and Fe chain was calculated as the total energy difference between their ferromagnetic (state "1") and antiferromagnetic (state "0") alignment, Fig. 4A (inset), within the broken symmetry approach. We find antiferromagnetic interaction between the Fe ladder and the Fe chain and the interaction strength decreases with distance in agreement with the experiment, (see Fig. 4A). To gain further insight into the long-distance limit (beyond 1.45 nm) of this interaction we approximated the Fe ladder by a Fe chain, which reduced the supercell size and enabled calculations up to 2.6 nm chain-chain separation. Remarkably, the decay trend of the magnetic interaction is comparable to the experiment: it is antiferromagnetic, non-oscillatory and decays with a characteristic length scale of approximately 1 nm, Fig. 4A. Some deviations from an approximate exponential trend are found for the smallest inter-chain distance. This, however, can be attributed to finite size-effects in the calculations (see Supplementary Materials for details).

We note that our DFT calculations do overestimate the magnitude of the interaction. This can be attributed to three effects: first, the employed exchange-correlation density functional (specifically the generalized gradient approximation) tends to delocalize the electronic wave functions; second, the small structural uncertainties in the vertical position of the Fe atoms above the $Cu_2N$; third, the lateral periodic boundary conditions that result in calculating Fe chains of infinite length instead of a finite nanobject. All three effects likely increase the calculated magnetic interaction energy compared to experiment. We remark that both the sign and decay behavior of the magnetic interaction are predicted accurately.

From our DFT calculations we can unambiguously link the long-rage magnetic interaction to the specific electronic properties of the surface. Hybridization between the Cu-$d$ and the N-$p$ orbitals in the $Cu_2N$ monolayer results in bonding and anti-bonding bands similar to bands reported for $Cu_3N$ films (31), (see Fig. 4B). The bonding bands are 6 eV below the Fermi energy, $E_F$, but the anti-bonding bands extend from approximately 1 eV below $E_F$ up to and across it. These bands are the source of the long-range magnetic interaction.

The anti-bonding Cu-N bands have either $\sigma$ or $\pi$ character. We find that the $\sigma$ band is mostly localized 1 eV below $E_F$, whereas the $\pi$ band extends over a wider energy range because of broadening induced by hybridization with the Cu substrate. The parallel Fe-atom chains give rise to an electronic confinement in the $Cu_2N$ layer that generates spin-polarized confined states stemming from these $\sigma$ and $\pi$ bands. The degree of confinement and the spin-polarization depend critically, and non-trivially, on the energy position of the Fe-$d$ orbitals with respect to the $\sigma$ and $\pi$ bands that the $d$-orbitals are allowed to hybridize with by symmetry. The resulting quantum well states are ultimately responsible for the interaction between the chains. This turns out to be antiferromagnetic and, most importantly, long range over a scale of 1 nm as illustrated in Fig. 4C, which displays the spin density profile and its decay in between two Fe atom chains.

To prove the key role the Cu-N network plays in setting the long-range magnetic interaction, we repeated the calculation breaking the Cu-N network by removing one row of N atoms between the two Fe atom chains. Now, the energy difference between the Néel states "1" and "0" drops below the numerical accuracy. In this case, the only way to achieve magnetic interaction is indeed via the Cu substrate, but such interaction is expected to be much smaller, short-range and ferromagnetic (29). Indeed, we were able to detect weak ferromagnetic interaction ($-12 \pm 3$ μeV at 2.1 nm distance) across two $Cu_2N$ patches experimentally by using a Fe trimer and a large 18-atom nano-antiferromagnet, Fig. 5.

It is worth noting that the magnetic interaction through $\sigma$ and $\pi$ states found for the $Cu_2N$ monolayer bears similarities to exchange coupling between the opposite edges of zig-zag graphene nanoribbons where confined states originating from the carbon $\pi$ band near the Fermi energy also cause exponentially decaying antiferromagnetic interaction (32, 33).

## Achievable sensitivity

The weakest magnetic interaction that is still detectable by the Fe trimer is determined by the smallest $\Delta T_1$ that can be measured above the noise of the pump-probe measurement. The pump probe measurements transduce $\Delta T_1$ into a variation of the number of electrons tunneling during the probe pulses so that the signal can be optimized by adjusting the probe pulse duration to be approximately equal to $T_1$. Increasing the pump-pulse amplitude or duration did not improve the signal but could induce additional switching of the nearby nano-antiferromagnets (see Fig. S2). We therefore kept the pump pulses minimal. The time traces measured with the Fe trimer at 3.0 nm distance from the nano-antiferromagnet have a signal amplitude of 0.55 e$^-$/pulse (electrons per probe pulse) and a noise level of 0.11 e$^-_{RMS}$/pulse for the chosen integration time of 0.1 s (Fig. 1c). This measurement indicates a magnetic interaction strength of 12 μeV and has a signal-to-noise ratio of 5. Hence, the sensitivity of this particular measurement was 0.9 μeV/√Hz which relates to an equivalent magnetic field sensitivity of 3.8 mT/√Hz. This sensitivity is sufficient that single spins could be detected at a distance of 2 nm exclusively via dipolar fields with 60 s integration time. This method surpasses the energy resolution of conventional scanning tunneling spectroscopy. This is possible because it measures the dynamics of the few-atom quantum magnet rather than static spectral features which are thermally broadened by 150 μeV for elastic and 230 μeV for inelastic tunneling spectroscopy at 0.5 K.

Fundamentally, the sensitivity of this spin sensing scheme is limited by the shot-noise of the tunneling electrons that probe the Fe trimer. Leaving all experimental parameters the same, a shot-noise limited detection would yield a sensitivity of 27 neV/√Hz (110 μT/√Hz). This limit has not been reached, but practical improvements of the sensitivity are possible by optimizing the applied bias field and tunnel junction setpoint (see Supplementary Materials Section 4 and Fig. S4).

Other nanoscale magnetometry methods using magnetic resonance force detection (4, 5), SQUIDs (6) or NV defects in diamond (7-10) achieve significantly higher sensitivities (tens of nT/√Hz). But the sensing objects in these methods are bulky, e.g. the cantilever for magnetic resonance force microscopy or the superconducting loop forming the SQUID, or must be embedded in an electrically insulating host, e.g. the NV center, is restricting them to sensing of dipolar stray fields at distances of tens of nanometers. Using electric read-out of a few-atom quantum magnet as demonstrated here is complementary in many aspects. Whereas it lacks in ultimate sensitivity it can be integrated onto metallic surfaces and in atomic proximity to sensed objects. Hence, it is compatible with atom manipulation techniques and sensitive to different interaction mechanisms, such as the Cu-N-mediated long-range indirect exchange interaction found here.

**Simultaneously sensing the magnetic states of multiple nano-antiferromagnets**

The sensitivity we achieved allows monitoring of several antiferromagnets within the detection range of one Fe trimer spin sensor. As a test we assembled one 12-atom and one 10-atom nano-antiferromagnet around a Fe trimer (Fig. 6A). Each has two Néel states giving four distinct configurations (labelled (0, 0), (0, 1), (1, 0) and (1, 1) in Fig. 6A). The intrinsic lifetime of each state is sufficiently long that we are able to identify each one of the four magnetic configurations and correlate each with a distinct $T_1$ of the Fe trimer by interleaving a series of pump-probe measurements with topographic imaging, Fig. 6B. Time traces of the pump-probe signal clearly detect four-state switching demonstrating that the Fe-trimer spin sensor can simultaneously detect the states of two nano-antiferromagnets, Fig. 6C.

Thereby we overcome a fundamental limitation of scanning tunneling microscopy measurements: the single-point measurement. Typically, information gained by the STM is local and limited to the single location of the STM tip. By using the Fe-trimer spin sensor we can simultaneously interrogate nano-objects at two different locations using a single probe, and perform non-local measurements that give access to spin-spin correlations. Such measurements are difficult to achieve by using conventional spin-polarized STM because scanning the probe tip over the nano-antiferromagnets would perturb their state population too much to reveal weak correlations.

We can test for such correlations with a histogram of the four-state switching signal measured for more than one hour (see Fig. S5). It reveals the probability, *P*, for each of the four states to occur, Fig. 6D. This probability distribution contains the probability with which the ferro- or antiferromagnetic configurations of the two nano-antiferromagnets occur. We find:

$$\frac{P_{(0,1)} + P_{(1,0)}}{P_{(0,0)} + P_{(1,1)}} = 1.12 \pm 0.09 \tag{2}$$

where ±0.09 is the standard error (±1σ) resulting from uncertainty in *P* due to the finite measurement duration of 4000 seconds. The value of 1.12 hints at a small but measurable antiferromagnetic correlation between the two nano-antiferromagnets.

## Conclusion

In conclusion, we have demonstrated that a few-atom spin system on surfaces can be used as an atomic-scale spin sensor to sense the magnetic state of single or multiple nano-antiferromagnets with micro-electron-volt energy sensitivity. In addition to surpassing the energy resolution of conventional STM spectroscopy the sensing scheme mitigates the dynamic and invasive influence of the tip on the sensed object. Here we applied the sensing scheme to quantify the distance-dependent magnetic interaction between a nano-antiferromagnet and a Fe trimer sensor on a $Cu_2N$ surface. The interaction was found to be long-ranged and antiferromagnetic. DFT calculations identify that the spin-dependent confinement of the $\sigma$ and $\pi$ bands of the $Cu_2N$ surface are the source for this antiferromagnetic interaction. This $Cu_2N$ substrate-mediated antiferromagnetic interaction presents an intriguing possibility to use artificial arrays of magnetic adatoms on $Cu_2N$ for studies of magnetic interaction across (quasi-) two dimensional $\pi$ electron systems and to explore some functionalities that have been originally proposed for 2D materials such as graphene nano-ribbons (33).

The sub-micro-electron-volt sensitivity of the sensing scheme enables indirect measurements of single-molecule magnets (36) that are difficult to address by direct tunneling spectroscopy and of ferromagnetic nanomagnets (34, 35) through their large dipolar stray field. These measurements do not require coherent control of the spin sensor and can in principle be implemented on other sample systems such as metallic (14), semiconducting (37) and thin insulating film (38) surfaces and with other few-atom sensors that feature two spin states at low energy. The ability to simultaneously sense the magnetic states of multiple nano-antiferromagnets enables atomic-scale studies of spin-spin correlations for classical and quantum-magnetic objects.

## Materials and methods

### Experimental design

This experiment aimed to demonstrate the possibility of creating integrated adatom spin sensors which allow spin-polarized STM to make highly sensitive, but non-invasive measurements of other adatom magnetic structures, including those with no net spin, on a surface. A series of experiments were devised to show an adatom chain coupled to a nano-antiferromagnet and to explore this coupling by tuning separation and magnetic field. An additional experiment was designed, placing the sensor in proximity with two nano-antiferromagnets, to show the potential for sensing multiple magnetic adatom structures.

All experiments were conducted using a low-temperature and ultrahigh-vacuum STM equipped with a 2 T vector magnetic field (Unisoku USM-1300 $^3$He). For all the measurements the temperature was maintained at 0.5 K, and the external magnetic field was aligned to the easy magnetic axis of Fe atoms in Fe trimer spin sensor. The easy axis was parallel to the direction of the two nitrogen atoms neighboring each Fe atom and the alignment accuracy of the magnetic field was ±3º accuracy.

The nano-antiferromagnet switched spontaneously between two Néel states (labelled "0" and "1") but at a sufficiently low rate that allowed topographic imaging (typical duration 20 s) and pump probe spectroscopies (60 s) to be acquired without a switch occurring during the acquisition. For measurements that required the nano-antiferromagnet to be in a particular state (Figs. 1B, 2B, 3C, 6B) we acquired a fast topography (10 s) before and after to verify that the nano-antiferromagnet did not switch. To attain statistical significance these measurements were repeated at least ten times.

**Sample preparation**

PtIr tips were sputtered with Argon and flashed by e-beam bombardment for ten seconds prior to use. The Cu(100) crystal was cleaned by several Ar-sputtering and annealing (850 K) cycles. After the last sputtering and annealing cycle that creates a clean Cu(100) surface, the monatomic copper nitride, $Cu_2N$, layer was prepared by nitrogen sputtering at 1 kV and annealing to 600 K. Then, the sample was precooled to 4 K and Fe atoms were deposited onto the cold sample by positioning it in a low flux of Fe vapor from a Knudsen cell evaporator.

The Fe trimer spin sensor was built by positioning Fe atoms 0.72 nm apart on the Cu binding sites of the $Cu_2N$ surface by vertical STM atom manipulation (13). The Fe trimer was built along the direction of the easy axis of Fe atoms in it. The nearby antiferromagnets were assembled from Fe atoms by the same technique. The spin-polarized tips were prepared by picking up 3 – 4 Fe atoms to the apex of the tip which yielded spin polarization $\eta \approx$ 0.1-0.3 (calibrated with 2 T external magnetic field). Spin polarized topographic images were acquired using a bias voltage below the 7.5 mV inelastic excitation thresholds of the trimer and nano-antiferromagnets to avoid inelastic excitation of either structure.

**All electronic pump-probe measurement**

An all-electronic pump-probe method was used to measure the spin relaxation time of the Fe trimer (18). A sequence of alternating pump and probe voltage pulses was created by a pulse pattern generator (Agilent 81110A) and sent to the sample using semi-rigid coaxial wires. The pump pulses excite the Fe trimer by inelastic scattering of tunneling electrons. The probe pulses detect the spin state of the Fe trimer because the tunnel magneto-resistance differs when the Fe trimer is in the excited state versus the ground state. Tunnel current resulting from the probe pulses was measured by lock-in detection at 690.6 Hz.

For the measurements shown in Fig. 2b, Fig. 3c and Fig. 6b the probe pulses were modulated on and off. This method removes the tunnel current contribution of the pump pulses from the lock-in signal. For the measurements shown in Fig. 1b, c, Fig. 5b and Fig. 6c the time delay between pump and probe pulses was modulated. This method removes any tunnel current contributions that are not due to time-dependent dynamics and records background-free time traces of the pump-probe signal (18). The average dynamical evolution of the Fe trimer was measured by slowly varying the time delay between pump and probe pulses, $\Delta t$. For increasing delay time the probability of Fe trimer still being in the excited state decreases exponentially. The spin relaxation time, $T_1$, was determined as the decay constant of an exponential decay function fitted to the delay-time dependent tunnel current $I(\Delta t)$. The typical time for taking the pump-probe spectrum (like Fig. 1b and Fig. 2b) and STM topograph is about two or three minutes.

**Density functional theory calculations**

All-electron DFT calculations were performed with the FHI-AIMS code (39). The Perdew-Burke-Ernzerhof (PBE) (40) generalized gradient approximation (GGA) to the exchange-correlation density functional was used. FHI-AIMS employs numerical atom-centered orbitals as basis set. Here, the "*light*" settings, which include the *tier 1* basis set for all atoms, was chosen. Nevertheless, for some selected cases, it was checked that energy differences changed by less than 1 meV per supercell when the *tier 2* basis set was used. This value (1 meV per supercell) was then fixed as limit to the numerical precision of total energy difference calculations and only results that satisfied this precision requirement are presented. Consistently, for each calculation, the convergence of the results with respect to the number of **k**-points was also carefully checked to ensure the same precision. Relativistic effects were included by means of the atomic zero-order regular approximation (41). The supercells considered for the calculations are described in the Supplementary Materials (Section S5). For the supercell with the smallest chain-chain (ladder) distance, the positions of all atoms were optimized until forces were smaller than 0.01 eV/Å. In all other calculations for larger chain-chain (ladder) distances, the relative coordinates of the Fe atoms and of its coordinating atoms are kept fixed to be the same and only the other atoms of the $Cu_2N$ surface and of the first two Cu layers were allowed to relax. By fixing the relative geometry of Fe coordination sphere, one is able to show that the decay of the magnetic interaction energy is due to the electronic structure and not to deviations in the Fe atom-to-surface distance.

# References

1. I. G. Rau, S. Baumann, S. Rusponi, F. Donati, S. Stepanow, L. Gragnaniello, J. Dreiser, C. Piamonteze, F. Nolting, S. Gangopadhyay, O. R. Albertini, R. M. Macfarlane, C. P. Lutz, B. A. Jones, P. Gambardella, A. J. Heinrich, H. Brune, Reaching the magnetic anisotropy limit of a 3*d* metal atom. *Science* **344**, 988–992 (2014).
2. L. Childress, M. V. Gurudev Dutt, J. M. Taylor, A. S. Zibrov, F. Jelezko, J. Wrachtrup, P. R. Hemmer, M. D. Lukin, Coherent dynamics of coupled electron and nuclear spin qubits in diamond. *Science* **314**, 281–285 (2006).
3. G. de Lange, Z. H. Wang, D. Ristà, V. V. Dobrovitski, R. Hanson, Universal dynamical decoupling of a single solid-state spin from a spin bath. *Science* **330**, 60–63 (2010).
4. C. L. Degen, M. Poggio, H. J. Mamin, C. T. Rettner, D. Rugar, Nanoscale magnetic resonance imaging. *Proc. Natl. Acad. Sci. USA* **106**, 1313–1317 (2009).
5. D. Rugar, R. Budakian, H. J. Mamin, B. W. Chui, Single spin detection by magnetic resonance force microscopy. *Nature* **430**, 329−332 (2004).
6. W. Wernsdorfer, From micro- to nano-SQUIDs: applications to nanomagnetism. *Supercond. Sci. Technol.* **22**, 064013 (2009).
7. G. Balasubramanian, I. Y. Chan, R. Kolesov, M. Al-Hmoud, J. Tisler, C. Shin, C. Kim, A. Wojcik, P R. Hemmer, A. Krueger, T. Hanke, A. Leitenstorfer, R. Bratschitsch, F. Jelezko, J. Wrachtrup, Nanoscale imaging magnetometry with diamond spins under ambient conditions. *Nature* **455**, 648–651 (2008).
8. J. R. Maze, P. L. Stanwix, J. S. Hodges, S. Hong, J. M. Taylor, P. Cappellaro, L. Jiang, M. V. Gurudev Dutt, E. Togan, A. S. Zibrov, A. Yacoby, R. L. Walsworth, M. D. Lukin, Nanoscale magnetic sensing with an individual electronic spin in diamond. *Nature* **455**, 644–647 (2008).
9. H. J. Mamin, M. Kim, M. H. Sherwood, C. T. Rettner, K. Ohno, D. D. Awschalom, D. Rugar, Nanoscale nuclear magnetic resonance with a nitrogen-vacancy spin sensor. *Science* **339**, 557–560 (2013).
10. M. Pelliccione, B. A. Myers, L. M. A. Pascal, A. Das, A. C. Bleszynski Jayich, Two-dimensional nanoscale imaging of Gadolinium spins via scanning probe relaxometry with a single spin in diamond. *Phys. Rev. Applied* **2**, 054014 (2014).
11. W. Paul, K. Yang, S. Baumann, N. Romming, T. Choi, C. P. Lutz, A. J. Heinrich, Control of the millisecond spin lifetime of an electrically probed atom. *Nat. Phys. Advanced Online Publication* doi:10.1038/nphys3965 (2016)
12. D. M. Eigler, E. K. Schweizer, Positioning single atoms with a scanning tunnelling microscope. *Nature* **344**, 524–526 (1990).
13. S. Loth, S. Baumann, C. P. Lutz, D. M. Eigler, A. J. Heinrich, Bistability in atomic-scale antiferromagnets. *Science* **335**, 196–199 (2012).
14. F. Meier, L. Zhou, J. Wiebe, R. Wiesendanger, Revealing magnetic interactions from single-atom magneticzation curves. *Science* **82**, 320 (2008).
15. A. A. Khajetoorians, J. Wiebe, B. Chilian, S. Lounis, S. Blugel, R. Wiesendanger, Atom-by-atom engineering and magnetometry of tailored nanomagnets. *Nat. Phys.* **8**, 497–503 (2012).

16. A. J. Heinrich, J. A. Gupta, C. P. Lutz, D. M. Eigler, Single-atom spin-flip spectroscopy. *Science* **306**, 466–469 (2004).
17. C. F. Hirjibehedin, C.-Y. Lin, A. F. Otte, M. Ternes, C. P. Lutz, B. A. Jones, A. J. Heinrich, Large magnetic anisotropy of a single atomic spin embedded in a surface molecular network. *Science* **317**, 1199–1203 (2007).
18. S. Loth, M. Etzkorn, C. P. Lutz, D. M. Eigler, A. J. Heinrich, Measurement of fast electron spin relaxation times with atomic resolution. *Science* **329**, 1628–1630 (2010).
19. S. Baumann, W. Paul, T. Choi, C. P. Lutz, A. Ardavan, A. J. Heinrich, Electron paramagnetic resonance of individual atoms on a surface. *Science* **350**, 417–420 (2015).
20. S. Yan, D.-J. Choi, J. A. J. Burgess, S. Rolf-Pissarczyk, S. Loth, Control of quantum nanomagnets by atomic exchange bias. *Nat. Nanotechnol.* **10**, 40–45 (2015).
21. X. Chen, Y.-S. Fu, S.-H. Ji, T. Zhang, P. Cheng, X.-C. Ma, X.-L. Zou, W.-H. Duan, J.-F. Jia, and Q.-K. Xue, Probing superexchange interaction in molecular magnets by spin-flip spectroscopy and microscopy. *Phys. Rev. Lett.* **101**, 197208 (2008).
22. A. A. Khajetoorians, M. Steinbrecher, M. Ternes, M. Bouhassoune, M. dos Santos Dias, S. Lounis, J. Wiebe, R. Wiesendanger, Tailoring the chiral magnetic interaction between two individual atoms. *Nat. Commun.* **7**, 10620 (2016).
23. L. Zhou, J. Wiebe, S. Lounis, E. Vedmedenko, F. Meier, S. Blugel, P. H. Dederichs, R. Wiesendanger, Strength and directionality of surface Ruderman-Kittel-Kasuya-Yosida interaction mapped on the atomic scale. *Nat. Phys.* **6**, 187–191 (2010).
24. V. Madhavan, W. Chen, T. Jamneala, M. F. Crommie, N. S. Wingreen, Tunneling into a single magnetic atom: spectroscopic evidence of the Kondo resonance. *Science* **280**, 567–569 (1998).
25. J. C. Oberg, M. R. Calvo, F. Delgado, M. Moro-Lagares, D. Serrate, D. Jacob, J. Fernandez-Rossier, C. F. Hirjibehedin, Control of single-spin magnetic anisotropy by exchange coupling. *Nat. Nanotechnol.* **9**, 64–68 (2014).
26. A. F. Otte, M. Ternes, K. von Bergmann, S. Loth, H. Brune, C. P. Lutz, C. F. Hirjibehedin, A. J. Heinrich, The role of magnetic anisotropy in the Kondo effect. *Nat. Phys.* **4**, 847–850 (2008).
27. A. Spinelli, M. Gerrits, R. Toskovic, B. Bryant, M. Ternes, A. F. Otte, Exploring the phase diagram of the two-impurity Kondo problem. *Nat. Commun.* **6**, 10046 (2015).
28. R. Pushpa, J. Cruz, B. Jones, Spin and exchange coupling for Ti embedded in a surface dipolar network. *Phys. Rev. B* **83**, 224416 (2011).
29. P. Wahl, P. Simon, L. Diekhöner, V. S. Stepanyuk, P. Bruno, M. A. Schneider, K. Kern, Exchange interaction between single magnetic atoms. *Phys. Rev. Lett.* **98**, 056601 (2007).
30. E. Simon, B. Újfalussy, B. Lazarovits, A. Szilva, L. Szunyogh, G. M. Stocks, Exchange interaction between magnetic adatoms on surfaces of noble metals. *Phys. Rev. B* **83**, 224416 (2011).
31. A. Navío, M. J. Capitán, J. Alvarez, F. Yndurain, R. Miranda, Intrinsic surface band bending in $Cu_3N$ (100) ultrathin films. *Phys. Rev. B* **76**, 085105 (2007).
32. H. Lee, Y.-W. Son, N. Park, S. Han, J. Yu, Magnetic ordering at the edges of graphitic fragments: Magnetic tail interactions between the edge-localized states. *Phys. Rev. B* **72**, 174431 (2005).
33. O. V. Yazyev, M. I. Katsnelson, Magnetic correlations at Graphene edges: basis for novel spintronics devices. *Phys. Rev. Lett.* **100**, 047209 (2008).

34. A. A. Khajetoorians, B. Baxevanis, C. Hübner, T. Schlenk, S. Krause, T. O. Wehling, S. Lounis, A. Lichtenstein, D. Pfannkuche, J. Wiebe, R. Wiesendanger, Current-driven spin dynamics of artificially constructed quantum magnets. *Science* **339**, 55–59 (2013).
35. A. Spinelli, B. Bryant, F. Delgado, J. Fernández-Rossier, A. F. Otte, Imaging of spin waves in atomically designed nanomagnets. *Nat. Mater.* **13**, 782–785 (2014).
36. M. Mannini, F. Pineider, P. Sainctavit, C. Danieli, E. Otero, C. Sciancalepore, A. M. Talarico, M.-A. Arrio, A. Cornia, D. Gatteschi, R. Sessoli, Magnetic memory of a single-molecule quantum magnet wired to a gold surface. *Nat. Mater.* **8**, 194–197 (2009).
37. A. A. Khajetoorians, B. Chilian, J. Wiebe, S. Schuwalow, F. Lechermann, R. Wiesendanger, Detecting excitation and magnetization of individual dopants in a semiconductor. *Nature* **467**, 1084–1087 (2010).
38. F. Donati, S. Rusponi, S. Stepanow, C. Wäckerlin, A. Singha, L. Persichetti, R. Baltic, K. Diller, F. Patthey, E. Fernandes, J. Dreiser, Ž. Šljivancanin, K. Kummer, C. Nistor, P. Gambardella, H. Brune, Magnetic remanence in single atoms. *Science* **352**, 318–321 (2016).
39. V. Blum. R. Gehrke, F. Hanke, P. Havu, X. Ren, K. Reuter, M. Scheffler, *Ab initio* molecular simulations with numeric atom-centered orbitals. *Comput. Phys. Commun.* **180**, 2175-2196 (2009).
40. J. P. Perdew, K. Burke, M. Ernzerhof, Generalized gradient approximation made simple, *Phys. Rev. Lett.* **77,** 3865 (1996).
41. E. Lenthe, E. J. Baerends, J. G. Snijders, Relativistic total energy using regular approximations, *J. Chem. Phys.* **101,** 9783 (1994).
42. N. Lorente, J. -P. Gauyacq, Efficient spin transitions in inelastic electron tunneling spectroscopy. *Phys. Rev. Lett.* **103**, 176601 (2009).
43. M. Ternes, Spin excitations and correlations in scanning tunneling spectroscopy. *New J. Phys.* **17**, 1–27 (2015).
44. S. Loth, C. P. Lutz, A. J. Heinrich, Spin-polarized spin excitation spectroscopy. *New J. Phys.* **12**, 125021 (2010).

**Acknowledgments:** The authors thank E. Simon for fruitful discussions and E. Weckert and H. Dosch, (Deutsches Elektronen-Synchrotron, Germany) for providing lab space. J.A.J.B. acknowledges postdoctoral fellowships from the Alexander von Humboldt foundation and the Natural Sciences and Engineering Research Council of Canada. A.D. and A.R. were supported by the European Research Council project QSpec-NewMat (ERC-2015-AdG-694097), the Marie Skłodowska-Curie project SPINMAN (SEP-210189940) and Grupos Consolidados UPV/EHU (IT578-13).

**Figure 1**

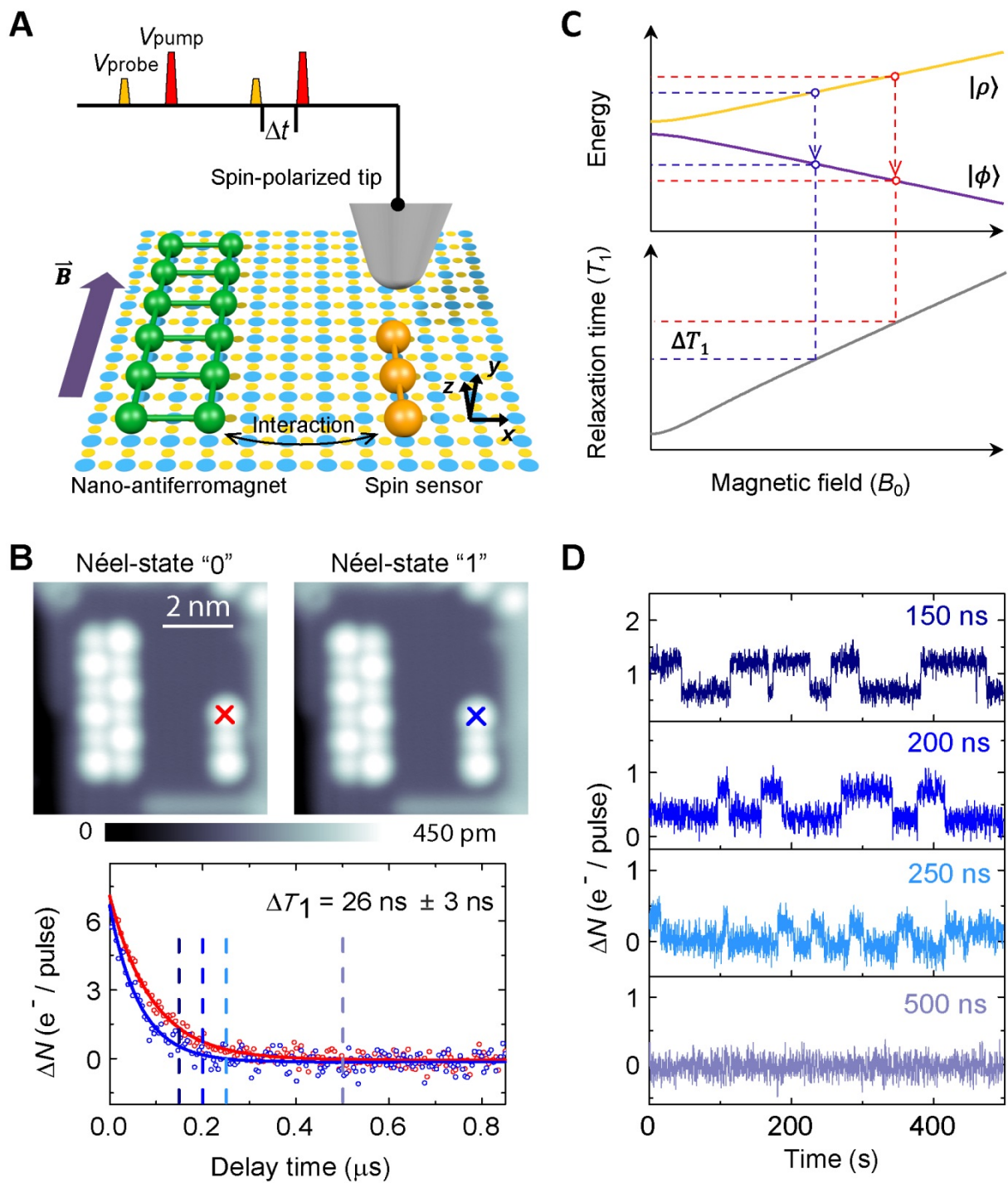

**Fig. 1. Sensing scheme with an atomic spin sensor.** (**A**) Schematic of the experimental setup. The Fe trimer spin sensor (orange) and a nano-antiferromagnet (light blue) are assembled from individual Fe atoms on a $Cu_2N/Cu(100)$ surface (Cu: yellow circles; N: grey circles) and interact weakly with each other. The spin-polarized probe tip of a STM (grey) is polarized by an external magnetic field (purple arrow). A series of pump and probe voltage pulses is sent to the tip and stroboscopically measures the spin relaxation time of the spin sensor. Coordinate system: z (easy magnetic axis of Fe

atoms in the Fe trimer), x (hard magnetic axis). (**B**) Top panel: spin-polarized STM topographs of the Fe trimer spin sensor and the Fe nano-antiferromagnet which switches between two Néel states (labelled "0" and "1"). The distance between Fe trimer and nano-antiferromagnet is 3.0 nm. Image size (6.6 × 6.6) nm$^2$, color from low (black) to high (white), tunnel junction setpoint, 5 mV, 50 pA. Bottom panel: pump-probe spectra of Fe trimer for the nano-antiferromagnet in Néel-state "0" (red dots) or "1" (blue dots). Tip position marked by cross in top panel. Pulse amplitude and duration: pump pulse, 8 mV, 80 ns; probe pulse: 3 mV, 100 ns. Solid lines are exponential decay fits to the experimental data showing that the spin relaxation time of Fe trimer differs by $\Delta T_1$ between the two curves. (**C**) Sketch of the avoided level crossing of the Fe trimer's low energy spin states, $|\phi\rangle$ and $|\rho\rangle$, that enables spin sensing. Changes in the magnetic field modify the energy splitting of the spin states and the transition rate between them (blue and red arrows) thus modifying $T_1$ by $\Delta T_1$. Any other magnetic perturbation that modifies the spin states also results in a $\Delta T_1$. (**D**) Time traces of the pump-probe signal measured on Fe trimer showing two-state switching of the nearby nano-antiferromagnet. The signal amplitude diminishes with increasing delay time between the pump and probe pulses (chosen delay times are indicated by vertical lines in (**B**)). Magnetic bias field 0.25 T and temperature 0.5 K.

**Figure 2**

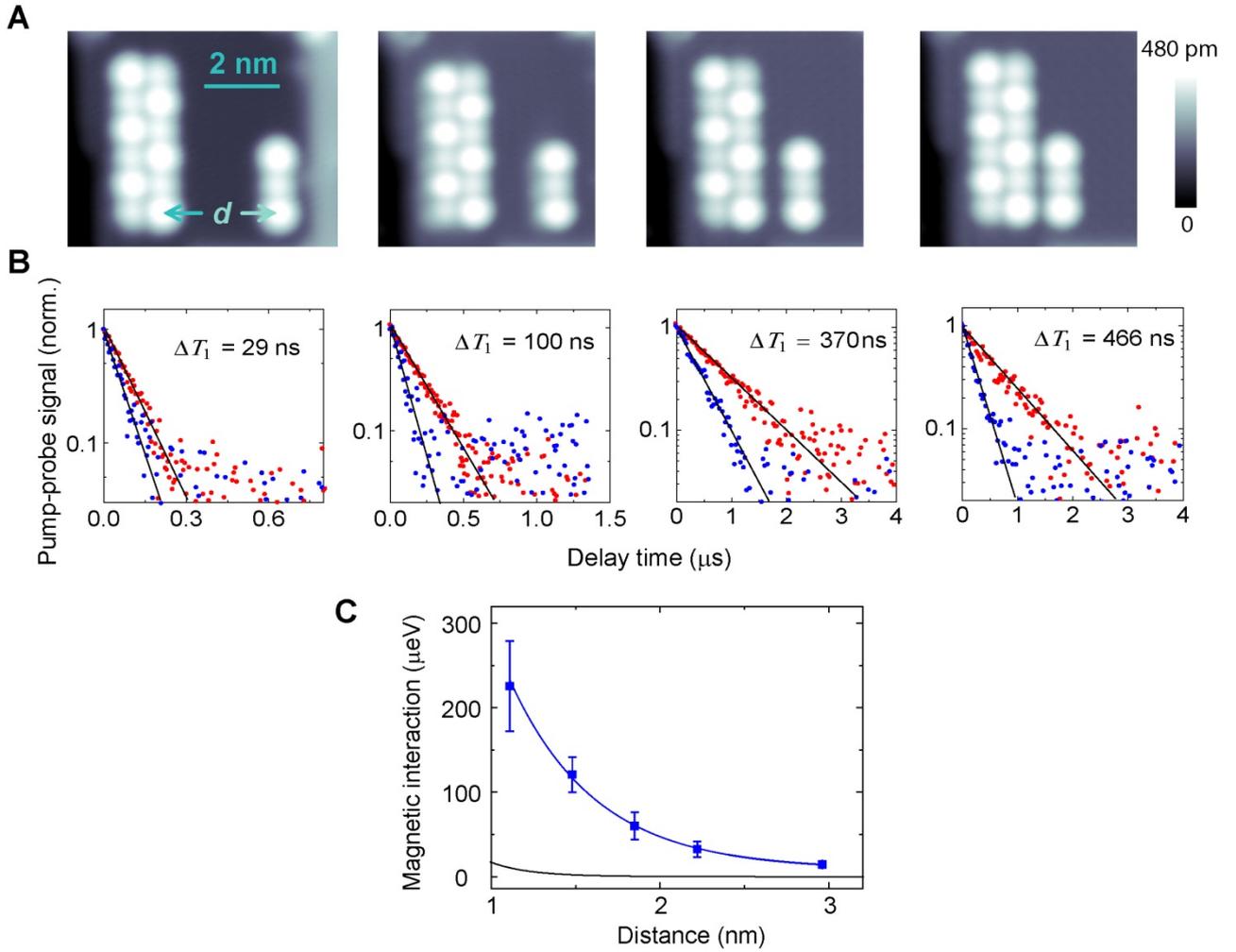

**Fig. 2. Distance dependence of nonlocal spin sensing.** (**A**) Spin-polarized STM topographs of different separations, *d*, between Fe trimer and nano-antiferromagnet. Tunnel junction setpoint, 5 mV, 20 pA. (**B**) Pump-probe measurements on the Fe trimer for each arrangement shown in (**A**) as the nano-antiferromagnet is in Néel state "0" (red dots) or "1" (blue dots). The pump-probe signal is normalized to 1 at zero delay time for clarity. Solid lines are exponential fits. Pulse amplitude and duration: pump pulse, 8 mV, 80 ns; probe pulse: 3 mV, 100 ns. Magnetic bias field during measurement 0.25 T, 0.5 T, 0.75 T, 0.75 T (left panel to right panel). (**C**) Magnetic interaction energy, *J*, between Fe trimer and nano-antiferromagnet for the arrangements shown (**A**) as a function of sensor-antiferromagnet separation (blue points). The blue line is an exponential fit to the measured interaction energies. The black line is the calculated magnetic dipolar interaction between Fe trimer and nano-antiferromagnet.

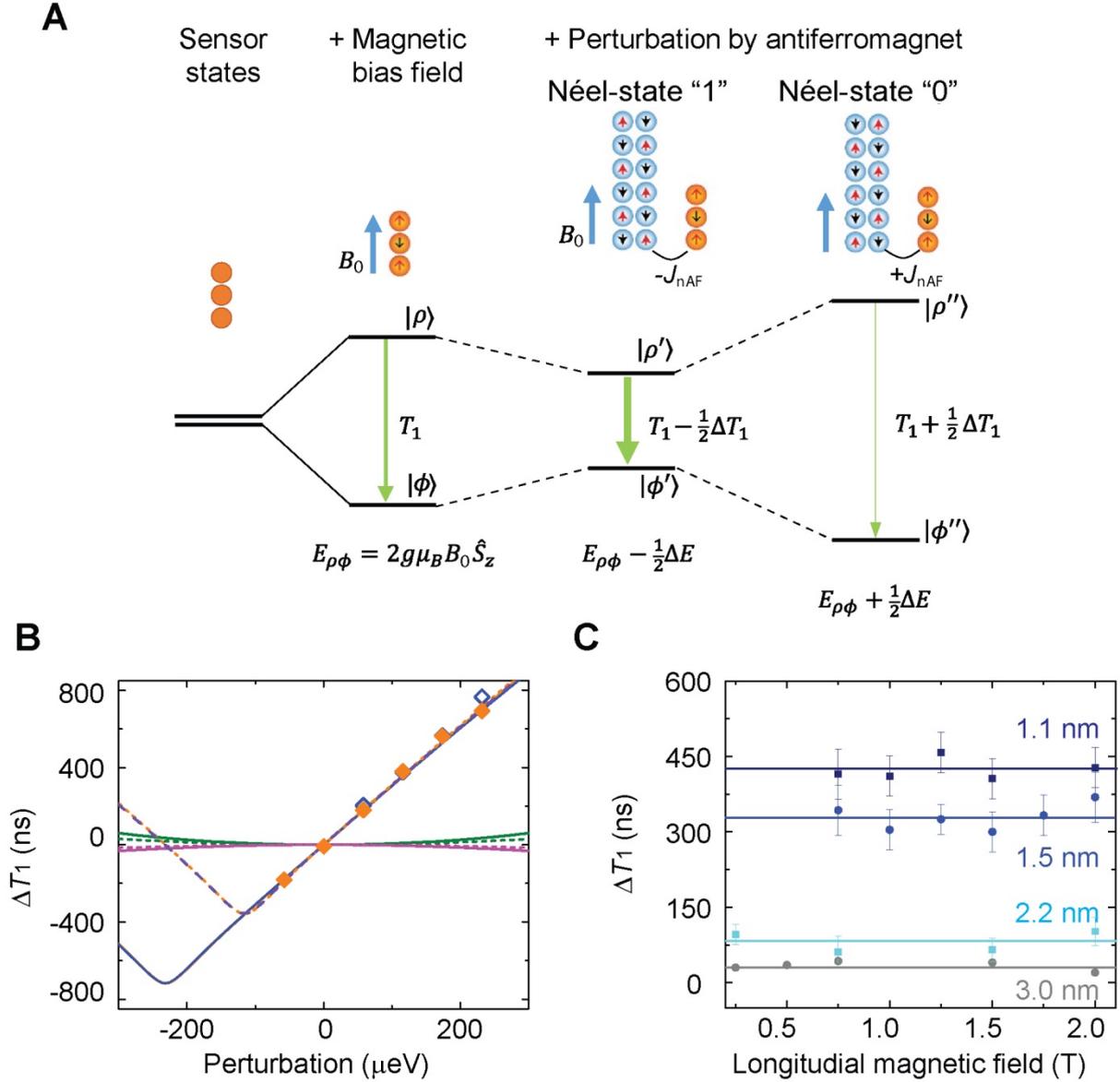

**Fig. 3. Sensing local magnetic interaction with a quantum spin sensor.** (**A**) Level diagram of the Fe trimer (orange atoms) showing the two low-energy spin states $|\phi\rangle$ and $|\rho\rangle$. In the absence of magnetic fields and other perturbations, the two spin states are very close in energy and spin relaxation time, $T_1$, is minimal. By applying an external magnetic bias field parallel to the $z$ axis, $B_o$ (blue arrow), the energy splitting increases to $E_{\rho\phi}$ given by the Zeeman energy. Magnetic interaction, $J_{nAF}$, with a nearby nano-antiferromagnet (blue atoms) adds a perturbation to the spin states that modifies $T_1$ by $\pm\Delta T_1/2$ and $E_{\rho\phi}$ by $\pm\Delta E/2$ depending on the Néel-state of the nano-antiferromagnet. (**B**) Variation of spin relaxation time, $\Delta T_1$, with magnetic perturbation, $\Delta E$, plotted for magnetic bias fields 1 T (solid lines) and 0.5 T (dotted lines). Calculated perturbations are longitudinal magnetic field parallel to the $z$ axis, $B_\parallel$ (orange), transverse magnetic fields, $B_{\perp x}$ (green), $B_{\perp y}$ (pink), and magnetic interaction with a nearby nano-antiferromagnet, $J_{nAF}$ (blue). The curves $\Delta T_1(B_\parallel)$ and $\Delta T_1(J_{nAF})$ are almost identical. Experimental data obtained by perturbing the sensor using an external

magnetic field, $B_\parallel$, and considering a bias field of $B_0 = 1$ T (orange dots) and 0.5 T (open dots). Plot shows Fe trimer at 1.5 nm distance from nano-antiferromagnet. (**C**) Difference in spin relaxation time of the Fe trimer, $\Delta T_1$, for the different Néel states of the nano-antiferromagnet plotted as a function of the longitudinal magnetic field for trimer-nano-antiferromagnet separation of 1.1 nm, 1.5 nm, 2.2 nm and 3.0 nm (dots). Experimental data (dots) and calculated behaviour (solid lines).

**Figure 4**

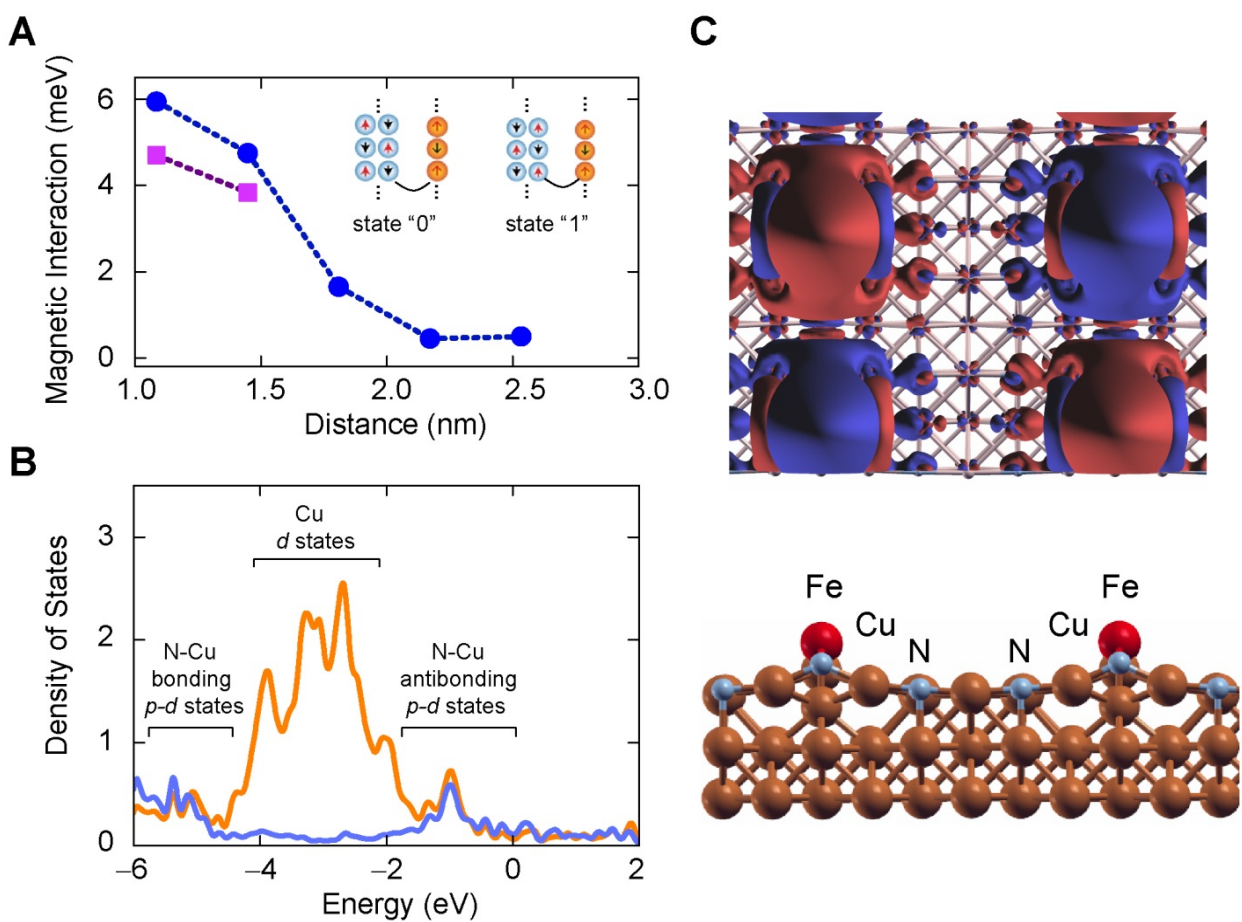

**Fig. 4. Calculated long-range magnetic interaction.** (**A**) Magnetic interaction energy between a Fe nano-antiferromagnet (blue atoms in inset) and a Fe chain sensor (orange atoms in inset) calculated by density functional theory, DFT, using broken symmetry approach (see Methods for details). Energies are calculated analogous to Fig. 3A. Within the DFT calculation the Fe-trimer sensor was approximated by an infinite one-dimensional chain of Fe atoms, and the nano-antiferromagnet by either an infinite ladder (purple dots) or another one-dimensional chain (blue dots). In the latter approximation a significantly smaller supercell could be used. This enables calculations for large distances but overestimates interaction energies due to absence of the second antiferromagnetically aligned chain in the nano-antiferromagnet (see Supplementary Figure S6 for supercell geometries). (**B**) Density of states of the $Cu_2N$ surface projected on the Cu atoms (orange) and N atoms (blue). Energies are given with respect to the Fermi energy. States between -4.5 eV and -1.5 eV are localized on Cu and have d-orbital character. States between -6 eV and -4.5 eV and between -1.5 eV and +0.5 eV are pd-hybrids between Cu and N of $\sigma$ and $\pi$ character. In particular the antibonding $\sigma$ bands give a sharp peak at $-1$ eV and the $\pi$ bands extend across the Fermi level. (**C**) Top panel: spin density distribution for two Fe atom chains at a 1.1 nm distance. Isosurfaces show $10^{-3}$ $e^-$/Å majority spin (red) and minority spin (blue) densities that extend well beyond the Fe atoms into the Cu-N network. Bottom panel: side view of the DFT supercell on same scale as top panel showing Fe atoms (red spheres), Cu atoms (orange spheres) and N atoms (blue spheres).

**Figure 5**

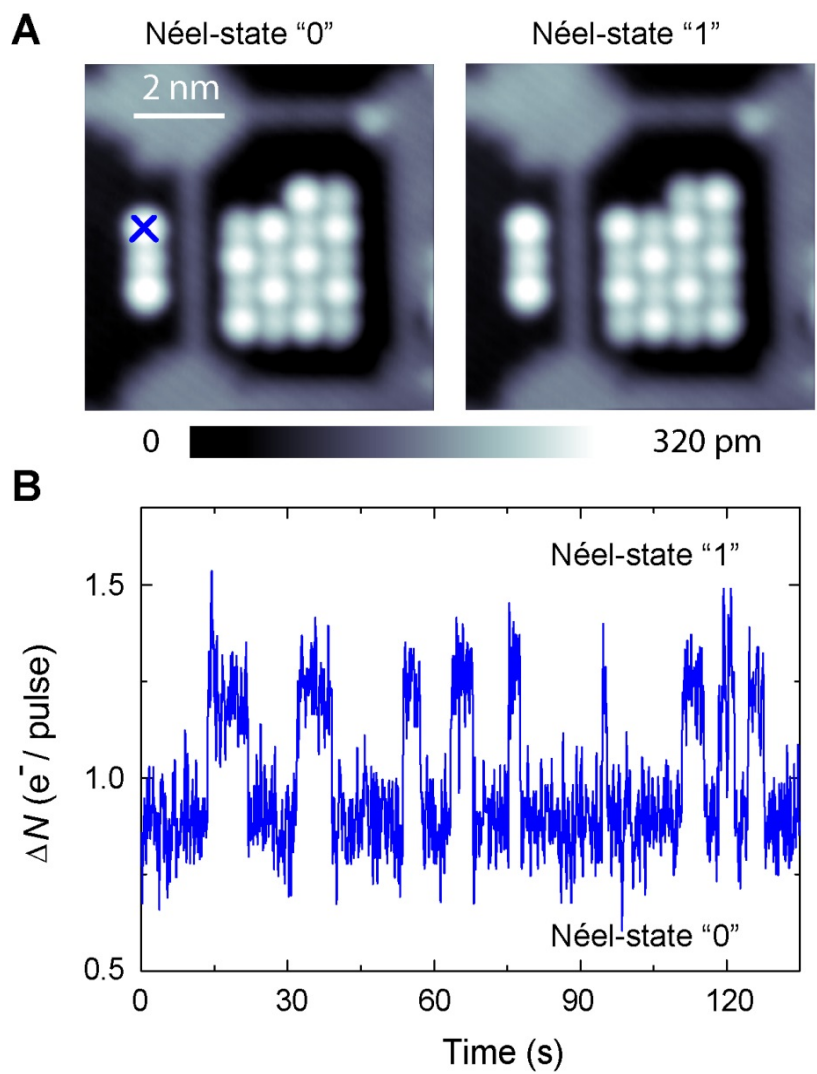

**Fig. 5. Sensing an nano-antiferromagnet located on different Cu₂N patch.** (**A**) Constant-current STM topographs of Fe trimer spin sensor and nano-antiferromagnet located on different Cu$_2$N patches. Image size, (7.7 × 7.7) nm$^2$. Tunnel junction setpoint, 5 mV, 10 pA. (**B**) Time trace of the pump-probe signal measured on Fe trimer resulting from Néel-state switching of the nano-antiferromagnet shown in (**A**), and the position of the STM tip during pump-probe measurement is shown as the blue cross. The measurement was taken with a 0.25 T external magnetic field. Pulse amplitude and duration: pump pulse, 10 mV, 50 ns; probe pulse: 3 mV, 100 ns; time delay between pump and probe pulse fixed at 150 ns.

**Figure 6**

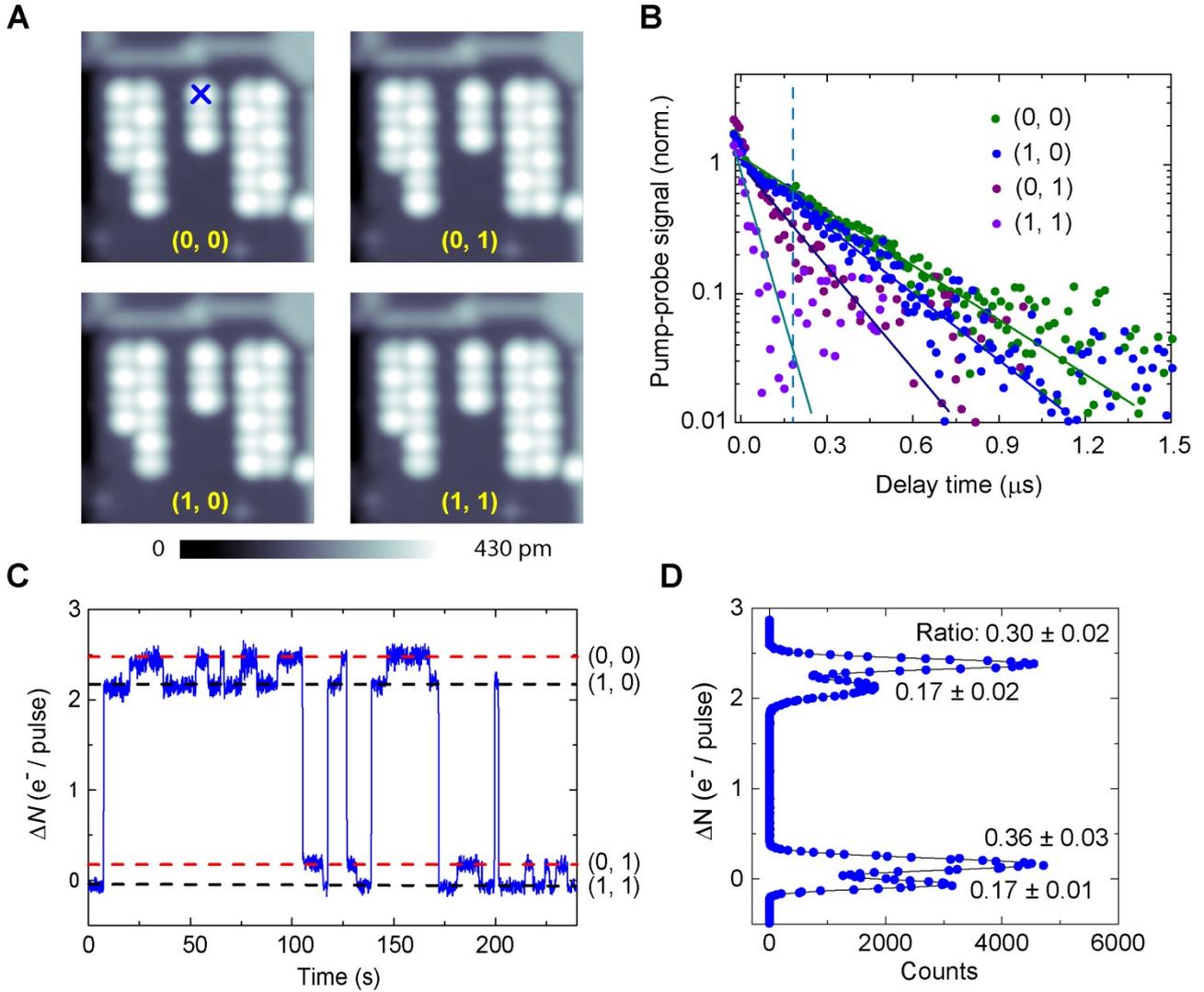

**Fig. 6. Simultaneously sensing the spin states of two Fe nano-antiferromagnets.** (**A**) STM constant current topographs showing two antiferromagnets constructed near a Fe trimer. The four magnetic configurations are labelled as (0, 1), (0, 1), (1, 0) and (1, 1), where the first (second) number indicates the magnetic state of the 10-atom nano-antiferromagnet (12-atom nano-antiferromagnet). Image size, (7.5 × 7.5) nm$^2$, tunnel junction setpoint, 5 mV and 20 pA. (**B**) Pump-probe spectra of Fe trimer (dots) recorded at the position marked in **a** for the four configurations in **a**. $\Delta T_1$ is 210 ±30 ns for a switch of the 12-atom nano-antiferromagnet and 60 ns ±10 ns for switching of the 10-atom nano-antiferromagnet indicating that the 12-atom nano-antiferromagnet interacts more strongly with the Fe trimer. Pulse amplitude and duration: pump pulse, 9mV, 50ns; probe pulse: 3mV, 100ns. (**C**) Time trace of the pump-probe signal measured on the Fe trimer (position marked by cross in (**A**)). The delay time between pump and probe pulses is 180 ns. Magnetic bias field, 1.5 T, tunnel junction setpoint, 5 mV, 500 pA. (**D**) Histogram of the state distribution shown in **c** but measured for 4000 seconds (see Fig. S5 for full time trace) with corresponding state occupation probabilities for the four configurations (0, 0), (1, 0), (0, 1) and (1, 1). The observed occupation probability may differ from the mean occupation probability due to the finite observation time. The stated error gives the standard deviation (±1σ) of the mean probability and was determined by resampling subsets of the measured time trace.